\documentclass[letterpaper, 10 pt, conference]{ieeeconf}  

\IEEEoverridecommandlockouts
\usepackage{graphics} 
\usepackage{epsfig} 
\usepackage{mathptmx} 
\usepackage{times} 
\usepackage{amsmath} 
\usepackage{amssymb} 
\usepackage{xcolor}
\usepackage[noadjust]{cite}
\usepackage{comment}
\usepackage{multirow}
\usepackage{array}
\usepackage{siunitx}
\usepackage{amsmath}
\usepackage{makecell}
\usepackage{graphicx}
\usepackage{lipsum}

\newcolumntype{P}[1]{>{\centering\arraybackslash}p{#1}}

\title{\LARGE \bf Improving Driver Situation Awareness Prediction using\\Human Visual Sensory and Memory Mechanism}

\author{Haibei Zhu, Teruhisa Misu, Sujitha Martin, Xingwei Wu and Kumar Akash
\thanks{The authors are with Honda Research Institute USA, Inc., San Jose, CA 95134, USA {\tt\small (hzhu, tmisu, smartin, xingwei\_wu, kakash)@honda-ri.com}}%
}

\begin{document}

\maketitle
\thispagestyle{empty}
\pagestyle{empty}

\begin{abstract}

Situation awareness (SA) is generally considered as the perception, understanding, and projection of objects' properties and positions. We believe if the system can sense drivers' SA, it can appropriately provide warnings for objects that drivers are not aware of. To investigate drivers' awareness, in this study, a human-subject experiment of driving simulation was conducted for data collection. While a previous predictive model for drivers' situation awareness utilized drivers' gaze movement only, this work utilizes object properties, characteristics of human visual sensory and memory mechanism. As a result, the proposed driver SA prediction model achieves over 70\% accuracy and outperforms the baselines.

\end{abstract}

\section{INTRODUCTION}

With improvements in sensing technologies, recent intelligent vehicles have been equipped with advanced driver assistance systems (ADASs) as standard features \cite{broggi2016intelligent, zhu2017overview}. For example, in the United States, forward collision warning system will be standardized to passenger vehicles by 2022 \cite{NHTSA2016a}. While those warnings have proven to reduce critical accidents, reports show that some users turn off these functions \cite{IIHS2016}. A major reason for this behavior is that those warnings are generated only based on surrounding traffic conditions and driver's steering and pedal {\it operations} but are oblivious to driver's {\it perceptions} and {\it decisions}. This results in making these warnings redundant as the system warns the driver even when the driver is already aware of the dangers.

To avoid these redundant warnings, the ADAS needs to be aware of drivers' awareness. In this work, we aim to sense and monitor the drivers' awareness status. While some studies and systems only use drivers' gaze movement as the proxy to driver's perception status \cite{kapitaniak2015application, topolvsek2016examination}, we tackle a more challenging problem of sensing drivers' memory status, which is required in accessing drivers' awareness. However, it is challenging to objectively assess if the driver is aware of a road hazard. One of the top challenges is that no reliable computational model for drivers' memory has been proposed that is able to generalize across various driving scenarios \cite{gugerty2011situation}.

The topic is refereed to as situational awareness (SA) \cite{endsley1988design, endsley1995toward}. It has been studied mainly in military applications, and recently, the concept has been applied to automobile applications \cite{endsley2016designing}. Eye-tracking is one of the viable options for SA estimation as it can be applied in real-time without interrupting the ongoing task \cite{moore2010development}. Furthermore, according to the eye-mind hypothesis \cite{just1980theory}, there is a close relationship between what the eyes are gazing at and what the mind is engaged with. However, the prediction performance of the previous results is not satisfactory because the previous study only considered aggregated gaze movement information \cite{kim2020toward}.

The contributions in this work are improvement of driver situational awareness prediction performance using the following approaches.
\begin{enumerate}
    \item We combined object properties (prior information of the objects) to driver gaze behavioral features that was used in the previous work.
    \item We extended gaze behavioral features considering the characteristics of human visual sensory systems. We demonstrated that drivers' gaze behavior with both foveal and peripheral visions was more effective in situational awareness prediction.
    \item We introduced a strategy to adjust driver awareness scores based on the psychology theory of human short-term memory capacity.
\end{enumerate}

The paper is organized as follows. We cover related work in Section II, followed by the data collection in Section III. Section IV describes the methods and features we utilized in prediction models. We analyze the prediction results of the proposed methods in Section V. We then clarify our research contributions through discussion in Section VI. Finally, we conclude our work in Section VII.

\section{RELATED WORK}


Recent advancements with machine-aided cognition and decision support systems have revealed the difficulty in collaboration between human and machine \cite{hancock2013human}. Situational awareness (SA) has been attracting increased interest from research activities \cite{endsley2016designing}. Endsley \cite{endsley1995toward} defined SA as the perception of environmental elements and events with respect to time or space, the comprehension of their meaning, and the projection of their future status. This concept has been applied widely to diverse applications, such as aviation applications, air traffic control, nuclear power plant, and vehicle operations \cite{moore2010development, nguyen2019review, gugerty2011situation}.

\begin{figure*}[!t]
    \centering
    \includegraphics[width=1.55\columnwidth]{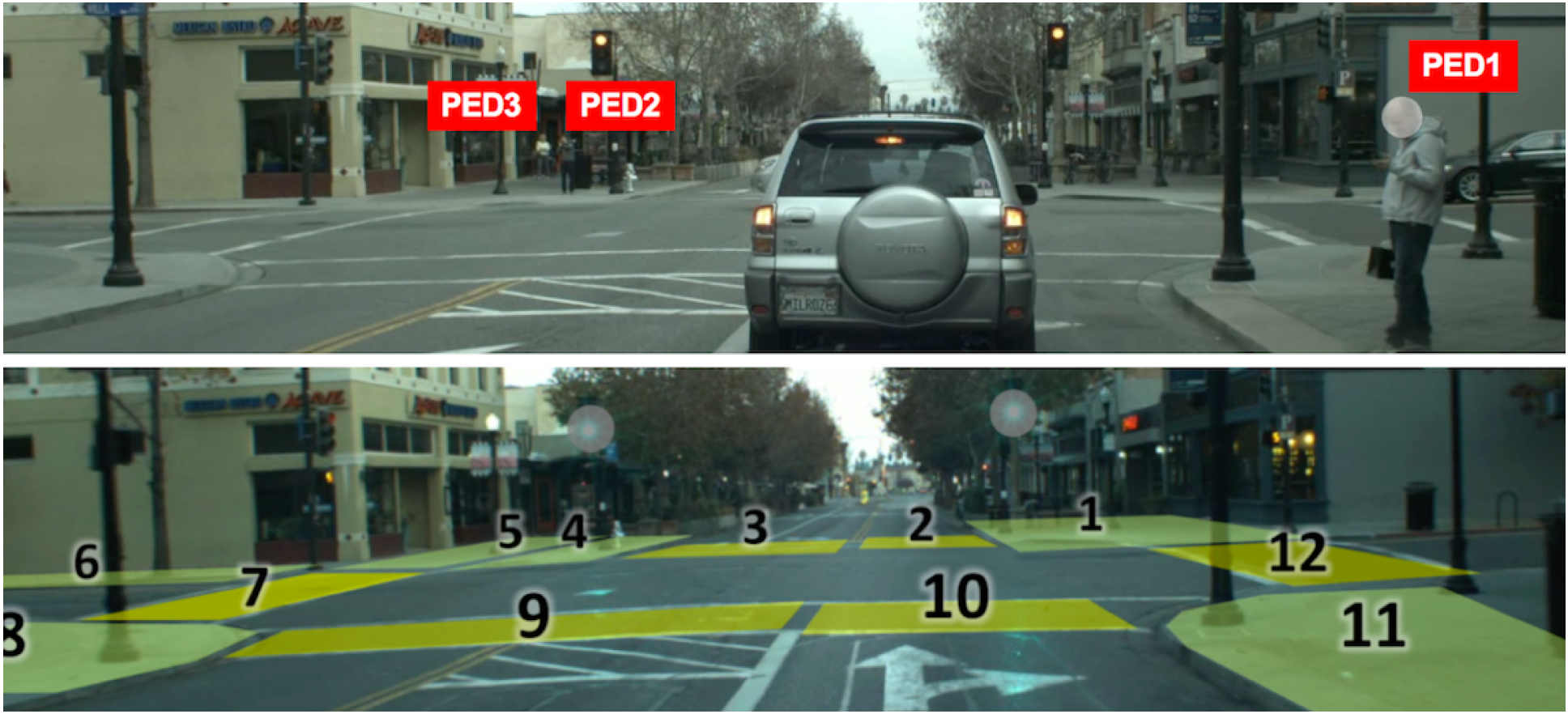}
    \caption{An example of a driving video (top, before a pause) and an empty scene (bottom, after the pause) presented on the wall projection screen to assess driver awareness of road hazards. The labels (e.g., PED1) are shown here for demonstration purposes and were not shown to the participants.}
    \label{fig:example_pause}
    \vspace{-3mm}
\end{figure*}

Numerous methods have been proposed to measure SA in human subject studies \cite{gugerty2011situation, kaber2012effects, hofbauer2020measuring}. Among these methods, the Situation Awareness Global Assessment Technique (SAGAT) \cite{endsley1988situation, endsley1995measurement, endsley2000direct} and the Situation Present Assessment Method (SPAM) \cite{durso1998situation} are commonly used. We used SAGAT to measure drivers' situation awareness in this study because of its effectiveness and efficiency. This offline query-based technique provides objective and direct measure of SA. In the SAGAT procedure, subjects are asked questions about a situation while the scenario is paused and the objects are hidden in the scene. Their answers with confidence levels are recorded. Since it's an offline approach, SAGAT cannot predict SA in real-time. The goal of this work is to estimate the SAGAT questionnaire result from gaze and object information towards real-time estimation of driver SA.

Previous studies have shown that SA is related to gaze behavior \cite{martin2018dynamics, kim2020toward}. However, further investigations of the relation between gaze behavior to moving objects of interest (OOIs), like vehicles and pedestrians, are necessary. We believe there are three major gaps that have not been considered in the previous studies. The first gap is the consideration of prior information about the object properties since different objects have different perceptual difficulties. For example, drivers are more likely to detect a pedestrian dressed in white comparing to others in dark. Driving context also impact the difficulty. For instance, drivers focus more on future trajectories, thus, traffic participants within those areas are likely to be looked at and aware of.

The second gap is that the previous study only considered gaze positions captured by the eye tracker \cite{lu2017much, sharma2016eye}. However, studies show that human uses not only their center vision but also peripheral vision to detect and track objects \cite{iwasaki1986relation, foley2019sensation}. The third gap is about the difficulty levels of perception and projection tasks. Existing studies have shown that the human short-term memory is limited \cite{miller1956magical, broadbent1975magic, chase1973perception}. Thus, we explored methods to approximate such short-term memory mechanism and consider the detection difficulty for objects. Also, the awareness to certain objects can be retained given their relative importance to other objects in the scene.

\section{DATA COLLECTION}

To estimate drivers' situation awareness and develop SA prediction models, a dataset collected by a human-subject experiment in a previous study was utilized \cite{kim2020toward}. This experiment utilized a video-based driving simulation scenario and SAGAT to collect drivers' SA answer. In general, all participants in this experiment went through the same procedure, in which they were asked to follow pre-defined routes and to answer questions about the situation of the intersection when the driving scene paused.

Each pause in experimental sessions presented a specific road intersection scene to participants (an example is shown in Figure \ref{fig:example_pause}). The top figure in Figure \ref{fig:example_pause} shows the last frame in the driving video before the pause, and the bottom figure shows the empty scene (i.e., the same scene as the scenario at the pause without any pedestrian or vehicle). Participants were asked to perceptually match locations of objects to the numbers presented in the bottom figure.

Since this was a driving simulation scenario, all participants were asked to control the steering wheel and pedals to mimic the driving prescribed in the video. Different types of data were collected from all sessions. Specifically, we focused on eye movement data \footnote{We performed automatic gaze registration as detailed in \cite{martin2018dynamics} to project the gaze onto the fixed scene perspective.} recorded by Tobii Pro Glasses 2 at 60 Hz and drivers' situation awareness answer collect by SAGAT \cite{kim2019assessing}.

In total, we had 44 participants, and each participant experienced 8 pauses with 28 different target objects, where participants' response corresponded to awareness or no awareness of each object. Thus, we have $44\times28=1232$ data points. Other than target objects, there are non-target objects in the scenes that the number of all objects are 7.5 on average with 3.0 standard deviation.

\section{FEATURES USED}

A 10 second analysis window has been used before each pause to capture the dynamic characteristics of objects and driver's perceptual engagement with said objects. We collected information, including objects' properties and participants' gaze movement, in this analysis window and aggregated them into object and gaze features.

\subsection{Gaze point-based features}

The first set of features is gaze point-based features, which utilized the positional relationship between target objects and participants' gaze center coordinates. These features were verified as highly correlated to drivers' awareness in a previous study \cite{kim2020toward}. Specifically, gaze coordinates were projected into the scene video, and ``gaze distance'' is defined as the distance between a gaze center and the nearest edge of the bounding box of a target object \footnote{We manually annotated the box of the position using VATIC \cite{vondrick2013efficiently}. All gaze-related features are calculated automatically based on the spatial relationship between projected gaze point and object bounding box.}. Three features measured in degrees are extracted as follows:

\begin{itemize}
    \item $G_{pause}$ - gaze distance to an object at a pause (right before SAGAT questions).
    \item $G_{min}$ - minimum gaze distance to an object within the analysis window before a pause.
    \item $G_{average}$ - average gaze distance of an object during an analysis window.
\end{itemize}

\subsection{Human visual sensory dependent features}\label{sec:fovea-feature}

Studies pointed out that while the sensitivity in object detection performance is low (less than 20\% capability in acuity compared to the foveal vision), the peripheral vision is good at the recognition of well-known structures and the detection of motions \cite{nelson1974motion}. In our task, we presume that drivers are using non-foveal vision for tracking detected objects and detecting moving objects. Based on the categorization of clinical research \cite{quinn2019clinical}, we used fovea ($\theta_1 = 2.5$ degree radius), perafovea ($\theta_2 = 4.1$ degree), perifovea ($\theta_3 = 9.1$ degree) and mocula ($\theta_4 = 15.0$ degree) and designed the following features. Specifically, the following three features are extracted per the above four radius ranges:

\begin{itemize}
    \item $HV^{\theta_n}_{elapse}$ - time length measured in second since the last frame where the gaze distance was less than a radius $\theta_n$ till the pause. A small value, for example less than 1 second, means the driver perceives the object right before the pause and is likely aware of it.
    \item $HV^{\theta_n}_{dwell}$ - total dwell time measured in second when the gaze distance was less than the radius $\theta_n$.
    \item $HV^{\theta_n}_{average}$ - average gaze distance in degree while the distance was within a radius $\theta_n$. If the gaze distance was always larger than this threshold, the average distance during the whole analysis window was used.
\end{itemize}

\subsection{Object spatial-based features}

Object properties have also been considered in our approaches since differences in objects' position information in a scene may impact drivers' awareness. Four features are extracted based on objects' annotated bounding boxes within the analysis time window:

\begin{itemize}
    \item $OS_{proximity}$ - distance between the object and the vanishing point of the scene, measured in degree.
    \item $OS_{duration}$ - accumulative time length that the object is visible in the scene within the analysis time window before a pause, measured in seconds.
    \item $OS_{size}$ - relative size of the object, also considered as the distance from the object to the ego-car approximated by the relative height of the object. Specifically, we normalized the height of the objects based on different object types, either vehicles or pedestrians.
    \item $OS_{density}$ - density of the scene, measured by the total number of objects in the scene.
\end{itemize}

\subsection{Object property-based features}

Other than objects' spatial-based features, property-based features also play an important role in accessing drivers' situation awareness. For example, a pedestrian crossing the driving lane would have a higher significance level and visual salience than a vehicle parking at the roadside regarding to the driver's view. For objects with high significance and salience, drivers could easily observe the object and take further actions to prevent potential accidents. Thus, the following six property-based features are extracted \footnote{Three annotators separately annotated those values based on an annotation guideline. In case of discrepancies they discussed to reach consensus.}:

\begin{itemize}
    \item $OP_{type}$ - object type, which is a binary feature representing if the object is a pedestrian or a vehicle.
    \item $OP_{relevance}$ - a binary indicator of whether an object's trajectory intersects with the ego-vehicle's trajectory within the analysis time window.
    \item $OP_{light}$ - traffic light condition for each object, representing green or red (or unknown) by a binary variable.
    \item $OP_{contrast}$ - object static salience or contrast against the background, representing the visual difference between the object and the background by an ordinal variable.
    \item $OP_{movement}$ - a manually annotated feature, representing the dynamic motion saliency of a target object at a pause. Each object is annotated as one of four categories: static (not moving), slow (regular walking speed), medium (regular vehicle speed), and high (fast vehicle speed).
    \item $OP_{change}$ - the area changing feature represents the potential projection errors of target objects by indicating if the object crosses the border of the target area within the last second before the pause. This feature is necessary because the driver may answer incorrectly because of such crossing trajectories.
\end{itemize}

\subsection{Human short-term memory-based features}

Previous studies did not consider human's short-term memory capacity. Physiology research such as Atkinson and Shiffrin's model \cite{atkinson1968human} suggested that human memory has three stores: a sensory register (1/4-1/2 seconds duration), short-term memory (0-18 seconds duration) and long-term memory (unlimited duration). We believe driver situational awareness is correlated with and significantly affected by short-term memory capacity.

\begin{figure}[!t]
    \centering
    \includegraphics[width=\columnwidth]{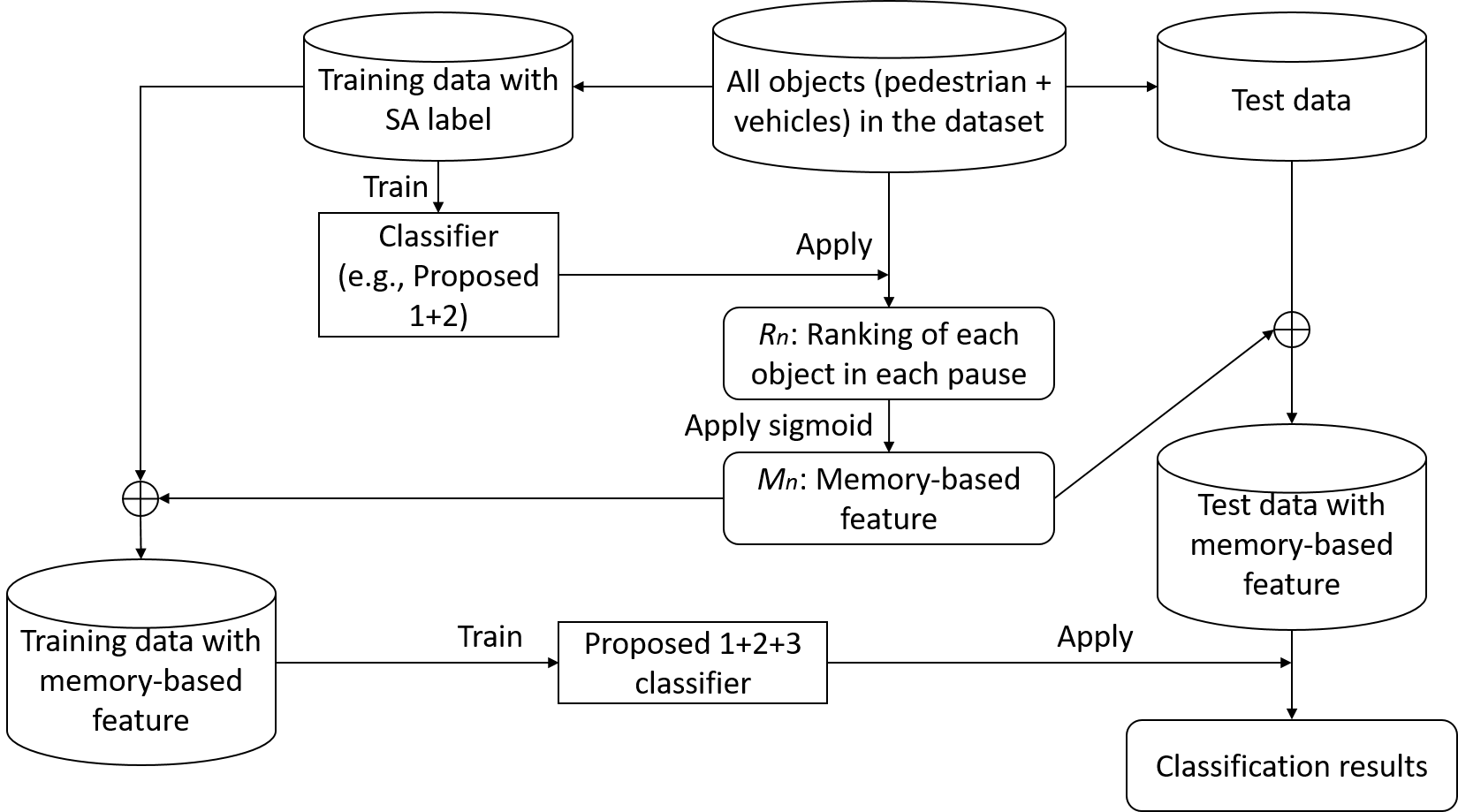}
    \caption{Flow of memory-based feature generation and its application for the classification task}
    \label{fig:propsed3}
\vspace{-3mm}
\end{figure}

We used one of the most famous finding in psychology -- The Magical Number Seven, Plus or Minus Two (a.k.a. Miller's law) \cite{miller1956magical} to approximate drivers' short-term memory capacity. This approach is expected to boost the probability of awareness of the top N objects and suppress others (even though they are salient). While the scene complexity feature implicitly incorporate this aspect, this feature boosts or suppresses score of all objects in the scene.

The SAGAT data only contains a single type (either vehicle or pedestrian) as target objects in a scene. However, a pause may include other non-target objects as well (e.g., Figure \ref{fig:example_pause} is a pause about pedestrian, but the scene has vehicles as well). And driver's memory for such non-target objects is not taken into account. Thus, this feature is designed and achieved by using a two-classifier structure, as listed as follows and illustrated in Figure \ref{fig:propsed3}.
\begin{enumerate}
    \item Apply a classifier trained based on all the features explained above and calculate SA scores for all objects in the dataset (i.e., both target and non-target \footnote{We calculated and annotated features for non-target objects in addition to target objects.} objects).
    \item Sort the objects in a scene based on the SA scores and get ranking $R_n$ for each object.
    \item The memory score for object $n$ ($M_n$) is calculated as
    \begin{equation}
        M_n = tanh(R_n - N)
    \end{equation}
    Here, N is a parameter that approximates the size of the human short-term memory.
\end{enumerate}
Note that the classifier for memory score calculation is trained based on the data in the training set and the method does not use the SA labels of the data in the test set.

\section{EXPERIMENTAL EVALUATION}

\subsection{Data preparation}

We evaluated the performance of our methods using the data collected by SAGAT. The data set consists of 8 pauses with 1,232 = (44 participants x 28 objects) data samples in total with SA labels. Due to the limited data size, we evaluated SA prediction performance using cross validation. We split the data into 8 portions based on 8 pauses that would be experienced by each participant (equals to one pause-out cross validation) because SA is strongly correlated at the object level but weakly correlated at the driver level \cite{kim2019toward}. Among these 8 pauses, we used the data from 7 pauses as the training data and the remaining pause as the test data.

We set class weights inversely proportional to class frequencies in the training data for our methods. Also, all features are normalized using MinMaxScaler so that each feature value ranges between 0 and 1. While using principal component analysis (PCA) for feature dimension reduction, we empirically set the number of PCA components to maximize the classification accuracy, which is the prediction success rate in testing datasets. Note that the chance rate of this classification task, where all samples are classified as negative (not aware) class, is 55.8 \%. For baseline and method evaluations, we focused on analyses on the operating characteristic (ROC) and the classification accuracy.

\begin{figure}[!t]
    \centering
    \includegraphics[width=\columnwidth]{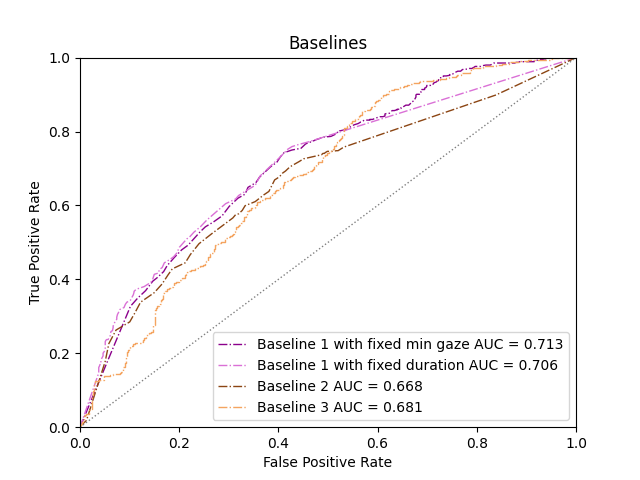}
    \caption{Evaluation baselines}
    \label{fig:baselines}
\vspace{-3mm}
\end{figure}

\subsection{Baseline Methods}

\subsubsection{Baseline 1: Rule-based baseline}

The primary assumption of this baseline is that drivers' eye-glance fixations directly equal to their situation awareness. We defined that the driver fixated on the object if the gaze stayed within 2.5 degree from the object for more than 120 milliseconds based on the finding that human can recognize information in fovea (2.5 deg \cite{quinn2019clinical, nelson1980functional}) within 120 ms \cite{rayner2007eye}. We draw ROC curves by changing 1) 2.5 degree to 0.1-30 degree range as well as 2) change 120 ms to 10-3000 ms range. As shown in Figure \ref{fig:baselines}, the area under the curve (AUC) values of these two ROC curves are  0.713 and 7.06. 

\subsubsection{Baseline 2: Learning-based baseline}

The fundamental assumption of this baseline is also that drivers' fixations equal to their situation awareness. Compared to Baseline 1, this baseline adopts a learning-based approach. With the human annotation of fixation based on the observation from gaze trajectory, we utilized a SVM model that predicts if the driver fixated on the object or not. The details of this approach are described in \cite{martin2018object}. We directly used the parameters in \cite{martin2018object} and did not fine tune the parameter based on our data. AUC value, which is obtained by changing SVM score (distance from the hyper plane), is 0.668 and the classification accuracy on testing datasets was 64.4\%.

\subsubsection{Baseline 3: Method based on the positional relationship between gaze center point and object positions}

This baseline uses gaze point-based features and object spatial-based features. However, human visual sensory dependent features, which is radius information as described in \ref{sec:fovea-feature}, are not used for this baseline. Rather than simply using these features, a PCA module with five components has been applied to the features to improve the baseline performance. The success rate in SA prediction by this baseline was 62.0 (we set classification threshold to maximize the success rate of the training set) and the AUC value was 0.681 as shown in Figure \ref{fig:baselines}.

\begin{figure}[!t]
    \centering
    \includegraphics[width=\columnwidth]{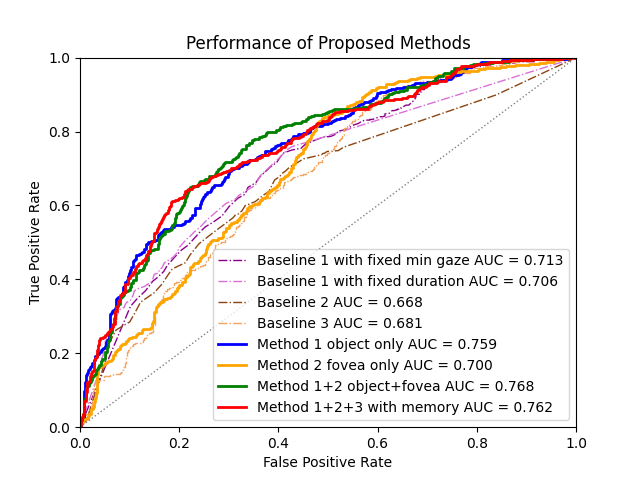}
    \caption{Model performance of Method 1, 2, and 3}
    \label{fig:performance}
\vspace{-3mm}
\end{figure}

\subsection{Proposed method 1: Effect of object property-based features}

The first method we proposed is to consider object property-based features into the estimation of drivers' SA. As mentioned in the Features Section, the main reason of considering these features is that we believe these properties could significantly effect the difficulty of detecting objects. Then, Method 1 was trained and tested with all Baseline 3 features plus object property-based features on a linear model of SVM. A PCA module has been applied to the features to reduced the feature dimension to 11 (PCA has also been applied in the following evaluations).

As shown in Figure \ref{fig:performance}, the performance of Method 1 is represented by the blue ROC curve, whose AUC value is 0.759. Clearly, such a performance outperforms all three baselines. This result indicates that object property-based features have impact on drivers' SA and can benefit SA prediction models.

\subsection{Proposed method 2: Effect of human visual sensory dependent features}

Understanding that different fovea ranges in human visual system have different characteristics of perceiving objects with varying properties, the second method we proposed is to consider human visual sensory dependent features (fovea-based features) by utilizing various characteristics of different fovea ranges of the human vision system. To evaluate proposed methods respectively, we first utilized the sensory-based features with features in Baseline 3 and developed a linear model of SVM for SA prediction. However, the performance improvement of the prediction model with such a feature combination is limited compared to Baseline 3, as shown in the ROC curves in Figure \ref{fig:performance} and the accuracy rates in Table \ref{tab:SA_accuracy}.

We then added object property-based features to the prediction model. In another word, we combined the sensory-based features with features used in Method 1. With all object and fovea features, the prediction performance outperforms all three baselines and Method 1, as the green ROC curve with AUC of $0.768$ shown in Figure \ref{fig:performance} and the prediction accuracy of $71.5\%$ listed in Table \ref{tab:SA_accuracy}.


\begin{table}[t]
\caption{SA prediction accuracy}
    \begin{center}
        \begin{tabular}{m{60mm}P{16mm}}
        \hline 
        \centering Method & Accuracy (\%) \\ \hline
        Chance rate (Percentage of majority class) & 55.8 \\ \hline
        Baseline 1 (Fixation rule) & 54.9 \\ \hline
        Baseline 2 (Fixation SVM)  & 64.4 \\ \hline
        Baseline 3 (Gaze+Object spatial feature)  & 62.0 \\ \hline
        Method 1 (Baseline 3+Object property features) & 69.9 \\ \hline
        Method 2 (Baseline 3+Sensory-based features)   & 62.3 \\ \hline
        Method 1+2 (Method 1+Sensory-based features) & 71.5 \\ \hline
        Method 1+2+3 (Method 1, 2+Memory feature)    & 72.4 \\
        \hline
        \end{tabular}
    \end{center}
\label{tab:SA_accuracy}
\vspace{-3mm}
\end{table}

\subsection{Proposed method 3: Effect of memory-oriented re-ranking features}

The third method we proposed is to approximate human memory system by implementing memory-oriented re-ranking features. As explained in the Feature Section, we developed a two-classifier model structure, in which the first classifier provides re-ranking features and the second classifier simulates the memory system by suppressing objects exceeding human short-term memory capacity and then estimates drivers' SA.

For the first classifier, we utilized both the proposed method 1 and 2, which contain all object and fovea features processed by PCA. Then, we applied the probability of each object of being aware of by the driver as part of the input for the second classifier. Specifically for the second classifier of a logistic regression model, we have tried multiple $N$ values as short-term memory capacity, and interestingly, the performance was the best when $N=7$ was used. This result aligns with the psychology finding of Miller's lab \cite{miller1956magical}.

\begin{table*}[ht]
\caption{An example of a detailed PCA analysis and interpretation -- coefficients of the top six PCA components with the highest contribution percentages (absolute values of coefficients over 0.2 were in bold font)}
    \centering
    \begin{tabular}{P{2mm}r|P{11mm}P{11mm}P{11mm}P{11mm}P{11mm}P{11mm}P{11mm}P{11mm}P{11mm}P{11mm}}
    \hline
    \multicolumn{2}{c}{\# Contribution} & $G_{pause}$ & $G_{min}$ & $G_{ave}$ & $OS_{proximity}$ & $OS_{duration}$ & $OP_{relevance}$ & $OP_{light}$ & $OS_{size}$ & $OS_{density}$ & $OP_{type}$ \\
    \hline
    1 & 25.7 & -0.04 & -0.06 & -0.02 & -0.09 & 0.09 & \textbf{0.26} & \textbf{0.39} & 0.09 & -0.02 & \textbf{-0.43} \\
    \hline
    2 & 17.2 & 0.09 & 0.16 & 0.07 & \textbf{0.23} & \textbf{-0.29} & 0.17 & 0.06 & -0.02 & 0.19 & -0.01 \\
    \hline
    3 & 11.7 & -0.04 & -0.08 & -0.03 & -0.12 & 0.18 & \textbf{0.42} & -0.10 & 0.11 & \textbf{-0.25} & 0.16 \\
    \hline
    4 & 10.0 & -0.03 & -0.11 & -0.04 & -0.08 & \textbf{0.22} & -0.18 & -0.02 & \textbf{-0.23} & \textbf{0.20} & \textbf{-0.25} \\
    \hline
    5 & 8.4 & -0.05 & -0.11 & -0.03 & -0.12 & -0.19 & \textbf{-0.24} & \textbf{0.35} & -0.10 & 0.16 & 0.16 \\
    \hline
    6 & 6.9 & 0.05 & -0.03 & 0.02 & 0.15 & \textbf{0.21} & \textbf{-0.35} & 0.19 & 0.17 & \textbf{-0.25} & \textbf{0.28} \\
    \hline
    & & $OP_{contrast}^{low}$ & $OP_{contrast}^{med}$ & $OP_{contrast}^{high}$ & $OP_{movement}^{static}$ & $OP_{movement}^{slow}$ & $OP_{movement}^{med}$ & $OP_{movement}^{high}$ & $OP_{change}$ & $HV^{\theta_n=2.5}_{elapse}$ & $HV^{\theta_n=2.5}_{dwell}$ \\
    \hline
    1 & & \textbf{-0.31} & -0.09 & \textbf{0.40} & \textbf{-0.31} & -0.03 & 0.13 & \textbf{0.21} & \textbf{0.29} & -0.07 & 0.09 \\
    \hline
    2 & & \textbf{-0.20} & \textbf{0.25} & -0.06 & 0.05 & \textbf{-0.45} & \textbf{0.25} & 0.15 & 0.18 & -0.03 & -0.12 \\
    \hline
    3 & & 0.18 & -0.12 & -0.05 & \textbf{0.43} & \textbf{-0.55} & 0.05 & 0.08 & 0.15 & 0.04 & 0.08 \\
    \hline
    4 & & -0.01 & \textbf{0.51} & \textbf{-0.49} & -0.05 & -0.08 & -0.01 & 0.14 & \textbf{0.29} & 0.13 & 0.01 \\
    \hline
    5 & & \textbf{0.20} & \textbf{-0.23} & 0.04 & 0.14 & -0.08 & \textbf{-0.47} & \textbf{0.41} & \textbf{0.25} & -0.01 & -0.13 \\
    \hline
    6 & & \textbf{-0.52} & \textbf{0.26} & \textbf{0.26} & \textbf{0.26} & -0.07 & -0.16 & -0.03 & -0.01 & 0.15 & -0.01 \\
    \hline
    & & $HV^{\theta_n=2.5}_{average}$ & $HV^{\theta_n=4.1}_{elapse}$ & $HV^{\theta_n=4.1}_{dwell}$ & $HV^{\theta_n=4.1}_{average}$ & $HV^{\theta_n=9.1}_{elapse}$ & $HV^{\theta_n=9.1}_{dwell}$ & $HV^{\theta_n=9.1}_{average}$ & $HV^{\theta_n=15.0}_{elapse}$ & $HV^{\theta_n=15.0}_{dwell}$ & $HV^{\theta_n=15.0}_{average}$ \\
    \hline
    1 & & -0.02 & -0.07 & 0.10 & -0.06 & -0.06 & 0.12 & -0.06 & -0.05 & 0.10 & -0.06 \\
    \hline
    2 & & 0.08 & 0.00 & -0.17 & \textbf{0.20} & 0.04 & \textbf{-0.28} & 0.17 & 0.09 & \textbf{-0.32} & 0.15 \\
    \hline
    3 & & -0.03 & 0.03 & 0.10 & -0.09 & 0.00 & 0.14 & -0.07 & -0.06 & 0.19 & -0.06 \\
    \hline
    4 & & -0.04 & 0.11 & 0.04 & -0.13 & 0.04 & 0.14 & -0.10 & -0.01 & \textbf{0.21} & -0.08 \\
    \hline
    5 & & -0.04 & -0.03 & -0.14 & -0.12 & -0.05 & -0.14 & -0.11 & -0.10 & -0.14 & -0.10 \\
    \hline
    6 & & 0.00 & 0.15 & -0.04 & -0.01 & 0.14 & -0.01 & -0.01 & 0.05 & 0.13 & 0.02 \\
    \hline
    \end{tabular}
    \label{tab:PCA}
\end{table*}

The performance of this method is represented by the red ROC curve in Figure \ref{fig:performance} with an AUC value of 0.762. Compared to the Method 1+2, the prediction accuracy rate improved from 71.5\% to 72.4\%, which was the highest among all baselines and propose methods. Also, we tested the proposed method 3 only (i.e., using Baseline 3 classifier to get memory-based features), but the performance of the classification did not improve. We believe the main reason is that the quality of the memory-based feature relies on the performance of the first classifier, and the quality of the ranking by the baseline 3 was not as good as the ranking by the proposed method 1+2 classifier.


\section{DISCUSSION}

To better understand how the features were used in proposed methods, we analyzed the meaning of PCA components \footnote{Note that this analysis is not for analysis of how the components affect to the classification.}. Table \ref{tab:PCA} summarizes the weights of all features for the top 6 PCA components with the highest contribution. We analyzed top 6 components because the cumulative explained variance ratio of the top 6 components were 80\%. 

It can be observed that weights for spatial-relationship based features ($OS_x$) are generally low. That means that these features are more noise rather than important information. Correlating with the result that the Proposed method 2 (using fixation information) outperforms Baseline 3, using human vision system-based knowledge are useful than using raw gaze movement information.

The first component has large weights with the features related to many of object property-based features including contrast, motion as well as the importance of the objects such as object type, relevance to ego-car future trajectory and traffic signal. We think this component summarizes features that represent the saliency (how easy the objects are to be detected) of objects. The second and third components has larger weight in total fixation time and movement, as well as state change of the objects. Interestingly, the weights of the second component for the wider field of view are larger. The weights for 2.5, 4.1, 9.1, 15.0 are -0.122, -0.173, -0.278 and -0.319. We believe these components characterize how people track (already detected) objects and people's peripheral vision has a relation to the tracking behavior.

In addition to the PCA modules, another important part in our methods is the two-classifier structure for approximating human short-term memory capacity. In our model, we used hyperbolic tangent function to model memory-based feature i.e., $M_n=tanh(R_n - N)$. We compared our definition with definitions using Heaviside step function (i.e., $M_n=1$ if $n>N$, $0$ if $n\leq N$) and linear function (i.e., $M_n=R_n$). We confirmed that the score using hyperbolic tangent was the best among those. We believe that the hyperbolic tangent outperforms the step function because it is more robust against the difference between the scenes and individual differences. The fact that our model outperforms the linear model indicates that the memory capacity is used equally for each objects rather than major objects take majority of the memory capacity.

The first major limitation of our work comes from the limited quantity of data that we were able to collect. Although our test scenes (road intersections) involve typical situations where most accidents happen \footnote{Approximately 40\% of accidents happen in intersections according to the Federal Highway Administration reports \cite{NHTSA2016b}}, it is unclear how our results can be generalized to other sites, traffic conditions, and weather conditions. Thus, the applicability of the proposed methods to other situations need to be explored further.

We are also aware of the limitation due to the data collection using a driving simulator. The simulator has a limited field of view of the traffic scenes, and its depth perception (due to flat projector screen) is disregarded. While we considered that participants in the data collection process were similar to the drivers in the real-world driving scenarios, their risk perception of traffic scenes could be different. Yet, we believe the proposed approaches and findings of this paper will provide essential and necessary steps towards developing human and situation-aware user assistance systems.

\section{CONCLUSION}

Understanding drivers' situation awareness can significantly benefit advanced driver assistance systems for intelligent vehicles in warning drivers about important traffic participants that could potentially cause accidents. In this work, we developed prediction models for estimating drivers' situation awareness by utilizing their gaze behavior, object spatial and property-based features, human visual sensory and short-term memory mechanism. The prediction performance of our models achieves over 70\% in accuracy rate and outperforms baselines established from previous studies.

While an increased fidelity of the experimental set-up as described in the limitation paragraph is an important future work, we would also like to test the effectiveness of our concept in the future. That is, we would like to implement an adaptive warning system by combining a real-time SA tracking system and object of importance estimation systems (such as \cite{martin2018object}) to locate important objects that drivers are not aware of and provide corresponding warnings. It is also an important future work to investigate the best strategy regarding when to make a warning or not to maximize driver's safety.

\addtolength{\textheight}{-0cm}   


\section*{ACKNOWLEDGMENT}

We gratefully acknowledge Dr. Hyungil Kim and Dr. Joe Gabbard for their assistance 
of data collection and annotation.

\bibliographystyle{ieeetran}
\bibliography{references}

\end{document}